\documentclass[twocolumn]{aastex631}
\shortauthors{Deesamutara et al.}

\usepackage{color}
\usepackage{amsmath}
\usepackage[]{natbib}
\usepackage{multirow}
\usepackage{graphicx}
\usepackage[caption=false]{subfig}

\graphicspath{{./}}

\begin{document}
    
\title{Extraction method for response functions from X-ray light curves of AGN by optimization algorithm}

\correspondingauthor{Tirawut Worrakitpoonpon}
\email{worraki@gmail.com}

\author[0000-0002-0964-0050]{Sanhanat Deesamutara}
\affiliation{School of Physics, Institute of Science, Suranaree University of Technology, Nakhon Ratchasima 30000, Thailand}

\author[0000-0002-0384-305X]{Tirawut Worrakitpoonpon}
\affiliation{School of Physics, Institute of Science, Suranaree University of Technology, Nakhon Ratchasima 30000, Thailand}
\affiliation{Center of Excellence in High Energy Physics and Astrophysics, Suranaree University of Technology, Nakhon Ratchasima 30000, Thailand}

\author[0000-0002-9099-4613]{Poemwai Chainakun}
\affiliation{School of Physics, Institute of Science, Suranaree University of Technology, Nakhon Ratchasima 30000, Thailand}
\affiliation{Center of Excellence in High Energy Physics and Astrophysics, Suranaree University of Technology, Nakhon Ratchasima 30000, Thailand}

\author[0000-0002-4516-6042]{Wasutep Luangtip}
\affiliation{Department of Physics, Faculty of Science, Srinakharinwirot University, Bangkok 10110, Thailand}
\affiliation{National Astronomical Research Institute of Thailand, Chiang Mai 50180, Thailand}

\author[0000-0002-9639-4352]{Jiachen Jiang}
\affiliation{Department of Physics, University of Warwick, Gibbet Hill Road, Coventry CV4 7AL, UK}

\author[0000-0002-6716-4179]{Francisco Pozo Nu{\~n}ez}
\affiliation{Astroinformatics, Heidelberg Institute for Theoretical Studies, Schloss-Wolfsbrunnenweg 35, D-69118 Heidelberg, Germany}

\author[0000-0003-3626-9151]{Andrew J. Young}
\affiliation{H.H. Wills Physics Laboratory, Tyndall Avenue, Bristol BS8 1TL, UK}

\begin{abstract}
We introduce a numerical optimization method to extract the X-ray reverberation response functions from the multi-band light curves of the active galactic nuclei. This approach does not require prior assumptions about the accretion disc and corona geometry, provided that the light curves result from the superposition of direct and singly-convolved signals, consistently across all bands. By reformulating the light curve equations into the matrix form, the optimal response matrix is derived by minimizing the squared difference between the observed and reconstructed light curves using a gradient-based optimization algorithm. We demonstrate that the method can robustly accommodate up to two convolution processes, such as the reverberation and the propagation, simultaneously.
When tested on the synthesized light curves, the method demonstrates robustness of the solutions to variations in the relative contributions of each light curve component as the recovered response kernel remains acceptably close to the ground truth, as evaluated by both the response geometry and the reconstructed light curves. The method’s tolerance to random noise was also assessed. With appropriate denoising, the response kernel can be reliably recovered when the signal-to-noise ratio is at least $100$. We show, as a proof of concept, that the proposed method is geometrical-model independent and has the potential to offer a flexible complement to traditional approaches.
\end{abstract}


\section{Introduction}
\label{sec:intro}

An active galaxy, a galaxy hosting an active galactic nucleus (AGN), has been a subject of interest because of its complexities as it consists of multiple components, each of which requires specialized theoretical and numerical treatments due to its distinct physical properties. An accretion disk, for instance, which is composed of charged hot gas and is situated close to the central supermassive black hole, is modeled in the gravito-magnetohydrodynamical formalism due to its dynamically active, highly agitated and charged natures (see \citealt{davis_et_al_2020araa} for the review). 
Overlying the accretion disk, the cloud of relativistic electrons, namely, the corona, is located. This component is a major subject of debates concerning the morphologies \citep{di_matteo_et_al_1997,galeev_et_al_1979, Cheng_2020, Liu_2023} and the thermo-optical properties \citep{fabian_et_al_2015}. The complexities of the studies of the AGNs are not only stemming from each individual component, but a lot of phenomena were found to emerge as a consequence of the interactions between them. For instance, the accretion of magnetized matter onto the black hole could produce the jet \citep{meier_2001,tchekhovskoy_et_al_2011,blanford_et_al_2019araa}. Such energetic emission affected the entire system including the corona \citep{markoff_et_al_2005,wang_et_al_2021}, the interstellar medium of the host galaxy \citep{antonuccio_delogu+silk_2008,olguin_inglesias_et_al_2016,su_et_al_2021}, or even the circumgalactic medium \citep{weinberger_et_al_2017,cielo_et_al_2018,martizzi_et_al_2019}. In addition, a component could also interact gravitationally with the other components, leading to the phenomena such as the disk precession  \citep{kumar+pringle_1985,nelson+papaloizou_2000,fragile+anninos_2005}, the coronal infall \citep{nakamura+osaki_1993,zycki_et_al_1995,xu_2015}, and the light bending \citep{bromley_et_al_1997,miniutti+fabian_2004,dovciak_et_al_2011}.

While the theoretical models and simulations of the AGN and its underlying physical processes has greatly been developed, the connection with the observations is possible mainly via the transmitted electromagnetic waves. As discussed above, the difficulty is that the electromagnetic waves emitted from a component can interact with the other components, causing the signals to be modified from their original forms. For instance, the optical/UV photons from multi-color blackbody radiation from the disk can undergo the inverse Compton scattering with the coronal electrons and get amplified to X-rays which changes significantly the emission spectrum \citep{Fabian2015}. On the other hand, the coronal X-rays can also be modified by reflecting on the disk before reaching the observer, so the disk composition can be drawn from the emission lines on the reflection spectrum \citep{Ross1999, Ross2005, Garcia2014, Garcia2020}. The reverberation time lag between the direct continuum and reflection X-rays have been used to determine the geometry of the disk-corona system \citep{McHardy2007, Fabian2009, DeMarco2013, Kara2016}. These factors render complexities and have to be taken into account in the analysis.

Focusing on the AGN X-ray light curve, the modification of coronal X-rays by any component is conventionally modeled by the convolutional formalism via the response functions depending on the geometry and the environment of the interacting component, such as in the phenomena of the reverberation from the accretion disc \citep{Uttley2014, Cackett2021}. Various models of corona geometries such as the lamppost model \citep{Wilkins2013, Emmanoulopoulos2014, Cackett2014, Chainakun2016, Epitropakis2016,Caballero-Garcia2018}, dual lamppost \citep{Chainakun2017, Hancock2023}, and radially extended corona \citep{Wilkins2016, Chainakun2019} have been proposed to understand the X-ray reverberation relating to geometries of the innermost region of AGN. However, in the conventional way, the study needs to assume the geometry of the AGN a priori, so that frame of work adheres strongly to the model chosen. Getting the information of the precise response function reflecting the true geometries of the system, is not straightforward. 

In this work, we propose the numerical method to extract the response functions underlying the generation of the multi-band light curves. We aim to resolve the limitation as mentioned before that the geometry of the corona, via a specific shape of the response function, has to be chosen before hand. We demonstrate that this approach is feasible under the assumption that the observed light curves are composed of superpositions of direct and singly-convolved driving signals, and that all energy bands share an identical reverberation process.
The multi-band light curves were also the starting point in the analysis of \citet{Reynolds_2000}, but the purpose and the methodology basically differ from ours. That work reconstructed an optimal light curve by a set of trial transfer functions, while we will directly optimize the response function for the temporal geometries that optimally reconstruct the target multi-band light-curves.

The article is organized as follows. First of all, the equations for the reverberated light curves, the mathematical formulation for the light curve equations is given in Sec. \ref{formula_response} before the numerical optimization method to solve for the response function is described in \ref{sol_response}. In Sec. \ref{sec:solutions} that follows, we test the method for different suites of light curves including those that are generated by a single reverberation response function and those that involve an additional propagation response function. Robustness of the method to many physical/numerical artifacts is also tested therein. Finally, we discuss and conclude our study in Sec. \ref{sec:conclu}.

\section{Light curve equations}
\label{formula_response}

Assuming a driving signal $a(t)$, the observed light curve $h(t)$ can be written as \citep{Emmanoulopoulos2014, Chainakun2016,Epitropakis2016}
\begin{equation}
 h(t) = b a(t) + R\int_0^t  \kappa (t-t^\prime) a(t^\prime)dt^\prime \;,
 \label{eq:lc}
\end{equation}
where $b$ and $R$ are the constants. The kernel $\kappa$ is the response function for any process that modifies a part of the driving signal. For instance, the X-ray reverberation from an accretion disc is a common process that is widely considered, and we adopt the response function $\kappa$ representing such process. The constants $b$ and $R$ can otherwise be regarded as the fraction of the light curve that passes directly to the observer and the other that is modified by the process whose response function is described by $\kappa$, respectively. This scenario is named the one-process case. 
Throughout the study, the parameter $b$ is the normalization factor tied to the continuum flux contribution, derived from the integral of the power-law spectral density as \citep{Emmanoulopoulos2014,Epitropakis2016}
\begin{equation}
    b = \int_{E_{\rm min}}^{E_{\rm max}} E^{-\Gamma}\,dE,
    \label{eq:b_energy}
\end{equation}
where $\Gamma = 2.0$. In case of the soft X-ray with $E_{\rm min} = 0.3\,{\rm keV}$ and $E_{\rm max} = 1.0\,{\rm keV}$, it yields $b=2.33$. The corresponding factor for the hard band, obtained from $E_{\rm min} = 1.5\,{\rm keV}$, and $E_{\rm max} = 10.0\,{\rm keV}$, is equal to $0.633$.
The prefactor $R$ is fixed to $1.0$ and $0.5$ for the soft and hard bands, respectively. A higher $R$ for the soft band is because it is reverberation-dominated.

More realistically, one can write the light curve equation with one additional term. We choose to include the long-timescale response function representing the accretion disk propagation \citep{Lyubarskii1997, Kotov2001, Arevalo2006}, and the light curve equation becomes
\begin{equation}
 h(t) = b a(t) + R\int_0^t  \kappa (t-t^\prime) a(t^\prime)dt^\prime + P\int_0^t  \pi (t-t^\prime) a(t^\prime)dt^\prime \;,
     \label{eq:lc_2p}
\end{equation}
which includes the second response function $\pi$ associated with the constant $P$. In this work, we employ the top-hat propagating function at the long timescale following~\cite{Alston2014} and \cite{Chainakun2023} for $\pi$.
We name the Eq. (\ref{eq:lc_2p}) the two-process light curve equation.
Unlike the one-process counterpart, the choices of all coefficients to generate the three-banded light curves are not based on the dependence on the band energy nor the physical nature of each band. The primary objective is to test the numerical feasibility when the two different kernels are solved simultaneously with an additional light curve band. As such, we set $b=1$ for all bands while $R$ and $P$ are chosen arbitrarily around the unity.
We choose the convention that the normalization of each response function is unity, so the partition between the direct, the reverberated, and the propagated signals can be adjusted by the prefactors before.

Input signals and response functions are generated in the similar way as mentioned in~\cite{deesamutara2025extractingxrayreverberationresponse}. The simulated driving signal is generated from a red noise power spectral density with index $\sim 2$ using the \texttt{stingray.simulator} package \citep{Huppenkothen2019}. The response functions are generated using the X-ray reverberation model \textsc{kynxilrev} \citep{Dovciak2004a, Dovciak2004b, Caballero2018}, which incorporates the \textsc{xillver} model for the reflection spectrum \citep{Garcia2010, Garcia2013}. 
In all cases, the driving signal is simulated with the black hole mass equal to $2 \times 10^6 M_{\odot}$   
~\citep{alston_dynamic_2020}, the inclination angle of $45^{\circ}$~\citep{Caballero-Garcia2020}, and the black hole spin parameter equal to $0.998$~\citep{Jiang2018}. These correspond to the physical parameters of IRAS 13224–3809.

Additionally, the Gaussian noise, derived from the Gaussian function $\mathcal{N}(\mu,\sigma)$, can be included into the signal. The variables $\mu$ and $\sigma$ in the Gaussian function designate the mean and the dispersion, respectively. The noise level relative to the signal can be evaluated by the signal-to-noise ratio (SNR) defined as 
\begin{equation}
    \text{SNR} =  \frac{\|\mathbf{s}\|^2_2}{\alpha^2 \|\mathbf{n}\|^2_2}  \;.
    \label{eq:s/n}
\end{equation} 
where $\|\mathbf{s}\|^2_2$ and $\|\mathbf{n}\|^2_2$ stand for the variances of the signal and the Gaussian function, respectively. For simplicity, we choose the zero-mean unit variance Gaussian function, namely, $\mathcal{N}(0,1)\equiv n$, so that the noise level is adjusted solely by the variable $\alpha$, as adopted by \citet{SNR_generator}.

Hence, the signal with the Gaussian noises included can be written in terms of $\alpha$ as,
\begin{equation}
         h(t) = b a(t) + R\int_0^t  \kappa (t-t^\prime) a(t^\prime)dt^\prime + \alpha n(t) \;,
    \label{eq:linear_lc_noise}
\end{equation}
and
\begin{equation}
 h(t) = b a(t) + R\int_0^t  \kappa (t-t^\prime) a(t^\prime)dt^\prime + P\int_0^t  \pi (t-t^\prime) a(t^\prime)dt^\prime + \alpha n(t) \;,
     \label{eq:linear_lc_noise_2}
\end{equation}
for the one- and the two-process light curves, respectively.

\section{The extraction of the response matrices by the optimization method}
\label{sol_response}

\begin{figure*}
    \centering
    \includegraphics[width=0.95\textwidth]{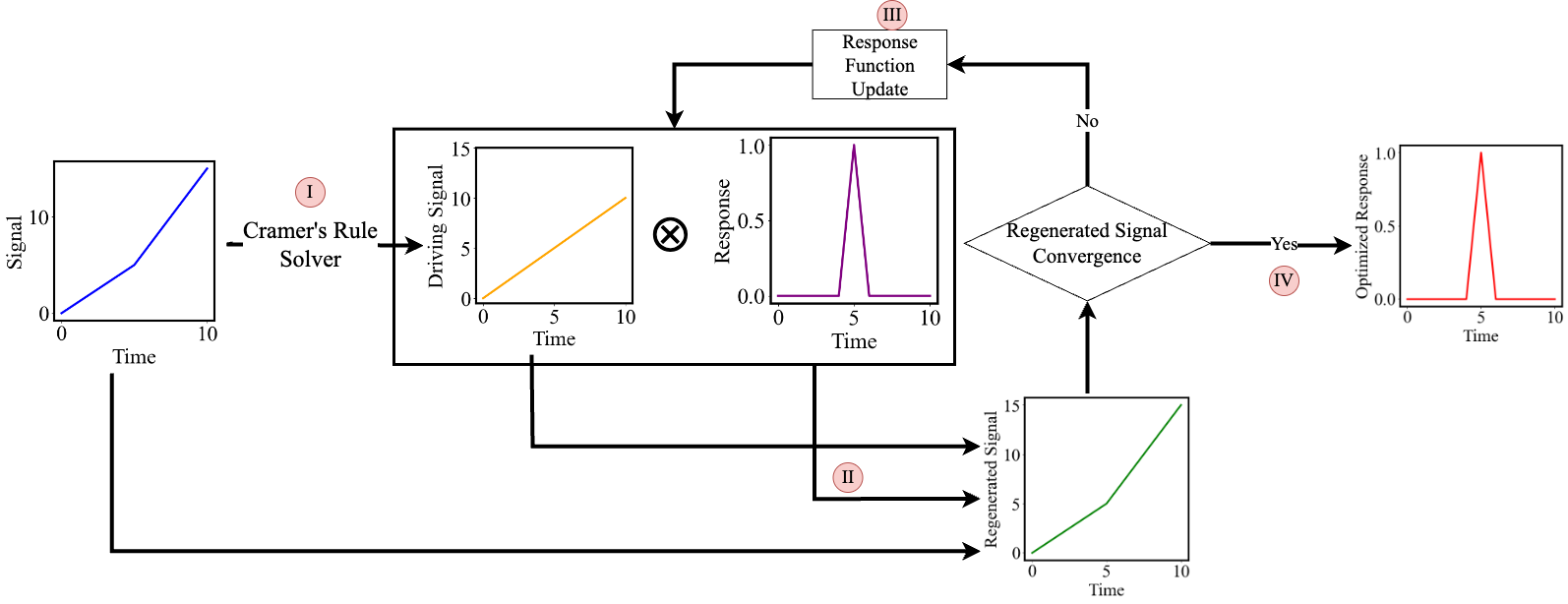}
    \caption{Workflow of the extraction method to solve for the response kernel. Once the targeted multi-band light curves ($h_s$ and $h_h$) are input, their driving signal, $a(t)$, is algebraically solved via the Cramer's rule (\textit{Step I}). After that, reconstructed light curves are generated by convolving $a(t)$ with the first guess of $\kappa(t)$ (\textit{Step II}). The similarity between the reconstructed and the observed light curves is then evaluated via the loss function. If the convergence has not been achieved, the response kernel is updated (\textit{Step III}) and the \textit{Step II} is repeated. Once the convergence is reached, the optimal response kernel is returned (\textit{Step IV}). It is important to note that the optimization focuses on minimizing the difference between the reconstructed and observed light curves, rather than between the reconstructed and ground-truth response kernels. This is because, in practical applications, the true response kernel is unknown—only the observed light curves are available as ground truth. See text for more details.}
    \label{fig:REV_Model}
\end{figure*}

The one-process light curve (\ref{eq:lc}) can be solved using the light curves from the two different bands. For instance, we can choose the soft-band $h_{s}$ and the hard-band light curves $h_{h}$ as inputs. Assuming that both bands share the same driving signal and their reverberation processes differ only by the prefactors, the light curve equations for the soft and and the hard bands can be expressed as
\begin{equation}
 h_{s}(t) = b_{s} a(t) + R_{s}\int_0^t  \kappa(t-t^\prime) a(t^\prime)dt^\prime \; ,
     \label{eq:lc_s}
\end{equation}
and
\begin{equation}
 h_{h}(t) = b_{h} a(t) + R_{h}\int_0^t  \kappa(t-t^\prime) a(t^\prime)dt^\prime \; ,
     \label{eq:lc_h}
\end{equation}
respectively. The principal assumptions in this study are: first, that the light curves across all energy bands originate from the same driving signal $a(t)$; and, second, that the reverberation response functions share the same shape but the different intensity of the reverberated signal is prescribed by the different prefactors before. 
To accompany the explanation of the method solving for the response function, the schematic workflow from the input light curves to the optimized response kernel in the one-process case is illustrated in Fig. \ref{fig:REV_Model}. 

Firstly, the light curve equations for the two bands, i.e., Eqs. (\ref{eq:lc_s}) and (\ref{eq:lc_h}), are reformulated into a generalized matrix form as follows:
\begin{equation}
    h_{l,i} = b_l a_i + R_l \sum_{j} \kappa_{ij} a_{j}
    \label{eq:linear_lc_matrix}
\end{equation}
where $l=s$ or $h$ stands for the band label. In this expression, $h_{l,i}$ and $a_{i}$ take the form of the $N$-row column matrix where $N$ is the dimension of the observed signal. We model the response kernel $\kappa$ as an $N\times N$ square matrix, i.e., $\kappa_{ij}$, where $i$ and $j$ represent the domains of $t$ and $t^\prime$, respectively. The fact that the response function is expressed as a function of $t-t^\prime$ for $t\geq t^\prime$ in the convolution term allows us to formulate the response matrix to be
\begin{equation}
\kappa_{ij}=\begin{cases}
\kappa_{i-j}, & \text{if $i \geq j$} \\
0, & \text{otherwise,}
\end{cases}
\label{eq:kappa_case}
\end{equation}
which can be illustrated as
\begin{equation}
    \kappa_{ij} =
    \begin{pmatrix}
             \kappa_0    & 0              & 0              & \dots     & 0 \\
             \kappa_1    & \kappa_0       & 0              & \dots     & 0 \\
             \kappa_2    & \kappa_1       & \kappa_0       & \dots     & 0 \\
             \vdots  & \vdots    & \vdots    & \ddots    & \vdots \\ 
             \kappa_{N-1}  & \kappa_{N-2}   & \kappa_{N-3}              & \dots     & \kappa_0
    \end{pmatrix}.
    \label{eq:kappa_mat}
\end{equation}
This definition yields the response matrix as a lower triangular matrix in which the values in the same right diagonal and sub-diagonal are identical. The matrix form of $\kappa$ in Eq. (\ref{eq:kappa_mat}) can be converted to the corresponding discrete function form, namely, $\kappa (t)$, by the series $\kappa_0$, $\kappa_1$, $\ldots$, $\kappa_{N-1}$. This reduces greatly the number of the degree of freedom of the response matrix from $N^{2}$ to $N$, which reduces greatly the numerical cost for the optimization.

If the driving signal and the response function are identical for the two bands, the two light curves constitute a set of linear equations with solvable driving signal and convoluted component if all prefactors are designated. We first of all compute the driving signal $a_{i}$ (\textit{Step I} in figure~\ref{fig:REV_Model}), whose solution is known to be
\begin{equation}
        a_i = \frac{h_{s,i}/R_s - h_{h,i}/R_h }{b_s/R_s - b_h/R_h}. 
    \label{eq:a_1p}
\end{equation}
Eq. (\ref{eq:a_1p}) suggests that the underlying driving signal is solvable for the next step if the direct-to-reverberation ratios for the two bands are not identical. This is a reasonable assumption as a lot of electromagnetic processes such as the radiative transfer, the absorption, or the reflection are known to be frequency-dependent.
Before we continue with the next step, it is worth clarified on what are the aim and the condition for the application of this workflow. It is true that in this paper, we test the method performance for light curves generated by known ingredients: the original driving signal, the underlying reverberation kernel, and the prefactors, so the \textit{Step I} to obtain the driving signal may appear unnecessary. This is not the case when tackling the real light curves because all of those ingredients are not know. For the workflow to be initiated, all prefactors have to be assumed. The solved driving signal and kernel thus represent the solutions specific to the choice of the prefactors, which are not necessarily the true ones. In other words, the optimization process can always be carried out for any valid combinations of the prefactors, but the closeness to the real one and the meaningfulness of the solutions should be of importance and investigated more in detail. These issues will be addressed in Sec. \ref{sec:solutions}.

If $b_{l}$ and $R_{l}$ are the guessed parameters, it is possible that a number of light curve bins fall below $0$. We additionally put a restriction that the solved $a_{i}$ is considered valid if less than $0.1N$ of the bins are negative. Then, the calculated negative bins are adjusted to zero. We do not further process the driving signal populated by a number of negative bins above that threshold. From the obtained $a_i$, the modeled light curve for the band $l$, namely, $\hat{h}_{l,i}$, is computed by
\begin{equation}
    \begin{split}
        \hat{h}_{l,i} = b_l a_i + R_l \sum_{j} \hat{\kappa}_{ij}a_j
    \end{split}
    \label{eq:linear_lc_iterative}
\end{equation}
where $\hat{\kappa}_{ij}$ is the modeled response matrix, corresponding to the guessed response matrix for the initial step. This corresponds to the \textit{Step II} in figure~\ref{fig:REV_Model}. We define the sum of the square of the difference between the real light curves $h_{l,i}$ and the modeled light curves $\hat{h}_{l,i}$ obtained from $\hat{\kappa}$, yielding the loss $\mathcal{L}_{1}$ that can be written as 
\begin{equation}
        \mathcal{L}_{1} = \frac{1}{N}\sum_l\sum_i \vert h_{l,i} - \hat{h}_{l,i} \vert ^2
    \label{eq:objective_1p}
\end{equation}
where the subscription $1$ denotes that this squared norm is for the one-process situation and it is used to update the response kernel toward a lower loss using the Adaptive Moment Estimation (ADAM) optimization method \citep{adamoptimizer} (\textit{Step III} in figure~\ref{fig:REV_Model}). The \textit{Steps II} and \textit{III} are repeated until the squared norm $\mathcal{L}_{1}$ is minimized, yielding the best-fitting $\kappa_{ij}$ specific to a chosen set of $b_{l}$ and $R_{l}$ (\textit{Step IV} in figure~\ref{fig:REV_Model}). As a proof of concept, the test of this optimization method to extract the convolutional response matrix in a simplified case in which the response function is a delta function is demonstrated in Appendix \ref{app:test_dirac_delta}.

The extractions of the two response kernels in the two-process light curve equation (\ref{eq:lc_2p}) require three different bands. If, as with the one-process counterpart, we assume the identical driving signal $a_i$ and two response functions $\kappa_{ij}$ and $\pi_{ij}$, the light curve equations from three different bands can be written in the matrix form as
\begin{equation}
         h_{l,i} = b_l a_i + R_l \sum_j \kappa_{ij}a_j + P_l \sum_j \pi_{ij}a_j \label{eq:linear_lc_iterative_2p_1} 
\end{equation}
where $l=1,2$ and $3$ refers to the band number. The matrix form of the response function $\pi_{ij}$ takes the similar form to $\kappa_{ij}$ as 
\begin{equation}
\pi_{ij}=\begin{cases}
\pi_{i-j}, & \text{if $i \geq j$} \\
0, & \text{otherwise.}
\end{cases}
\label{eq:pi_case}
\end{equation}
The driving signal $a_{i}$ can be solved by the Cramer's rule and the solution reads
\begin{equation}
    a_i = \frac{   \det \begin{pmatrix}
            h_{1,i}  &  R_1  &  P_1  \\
            h_{2,i}  &  R_2  &  P_2  \\
            h_{3,i}  &  R_3  &  P_3       
    \end{pmatrix}}
    {\det \begin{pmatrix}
            b_1  &  R_1  &  P_1  \\
            b_2  &  R_2  &  P_2  \\
            b_3  &  R_3  &  P_3       
    \end{pmatrix}},
    \label{eq:a_2p}
\end{equation}
and $a_{i}$ is considered valid if the number of negative bins is less than $0.1N$, as for the one-process case. Note that the Cramer's rule permits the solutions for both of the convoluted components which will be used for the loss function in the optimization process. The negative bins of the solved driving signal are treated in a similar way as for the one-process case. Then, the modeled two-process light curve, in the matrix form, for each band is given by
\begin{equation}
        \hat{h}_{l,i} = b_l a_i + R_l \sum_{j} \hat{\kappa}_{ij}a_j + P_l \sum_{j} \hat{\pi}_{ij}a_j,
    \label{eq:linear_lc_iterative_2p}
\end{equation}
where $\hat{\kappa}$ and $\hat{\pi}$ are the modeled response matrices. The solutions for the two modeled response matrices are determined by minimizing the squared norm $\mathcal{L}_{2}$ defined as
\begin{equation}
        \mathcal{L}_{2} = \frac{1}{N}\sum_i \biggl( \vert (\kappa \otimes a)_{i} - \sum_{j} \hat{\kappa}_{ij}a_j \vert ^2 + \vert (\pi \otimes a)_{i} - \sum_{j} \hat{\pi}_{ij}a_j \vert ^2\biggr).
        \label{eq:objective_2p}
\end{equation}
We choose a different loss function from the one-process case as in this case, the loss is calculated between the convoluted terms from the Cramer's rule and the convoluted terms from $\hat{\kappa}_{ij}$ and $\hat{\pi}_{ij}$, which are on track of the optimization, instead of determining it from the full light curves to reduce the computation time. We will demonstrate that this choice of the loss function yields the comparable robustness to that for the one-process counterpart which is determined from the full light curves.

The numerical optimization source code is developed with PyTorch library \citep{Pytorch}, with objective functions defined as Eqs. (\ref{eq:objective_1p}) and (\ref{eq:objective_2p}) for the one-process and the two-process cases, respectively. In principle, this process leads $\hat{\kappa}$ and/or $\hat{\pi}$ to the optimal ones at the end. The starting point of the response kernels is the two-leveled step function of area unity with the cutoff at $t=1000$ s. The workflow's optimization rate (equivalent to the learning rate in machine learning) has to be low, i.e., $1.0 \times 10^{-5}$, as the pipeline itself suggests. A high optimization rate leads to an oscillating loss around the optimal loss without convergence. The computation time can alternatively be reduced by appointing the adaptive learning rate scheduler. Applying this lets the optimization rate decline along the epochs by a certain prescription such as linear or exponential.
Solved response kernels can be further cross-matched with the library of the theoretical response kernels of various scale heights, simulated from the ray tracing. Determination is done under eighty-percent rule, following method mentioned in Section 6.2 of \cite{deesamutara2025extractingxrayreverberationresponse}.

Solving for the response function by this method differs from the conventional approaches as, first of all, it is not necessary to step into the frequency domain, thus the numerical artifacts from the conversion can be avoided. Only a suite of simultaneously observed multi-band light curves is required as input, which are acquirable from various X-ray space telescopes, e.g., the \textit{XMM-Newton}, and the \textit{NuSTAR}. 
Secondly, it is not required to specify the model of the response function as the starting point. Thus, the response kernel obtained from the multi-band light curves corresponds to the optimized kernel specific to choices of the prefactors of all bands. We expect that the optimized kernels from the light curves with prefactors being exactly those used for the generation of the light curves should coincide with the true kernels. The tests of the robustness will be carried out in the following section.

Multi-band light curves were also analysed by \citet{Reynolds_2000}, however, the goal and methodology differ from ours. In that work, the continuum light curve is first optimally reconstructed and then folded through a set of trial transfer functions, and the existence of a lag is assessed by identifying the delay parameters that minimize the $\chi^{2}$ difference between the modeled and observed line-band light curves. The emphasis is therefore on testing for characteristic time delays under assumed response forms. In contrast, our approach directly optimizes the response function itself, aiming to recover its temporal geometry by jointly reconstructing multi-band light curves from a common driving signal.

\section{Results}
\label{sec:solutions}

\begin{figure*}
  \centering
  \begin{tabular}{ccc}
    \includegraphics[width=0.3\textwidth]{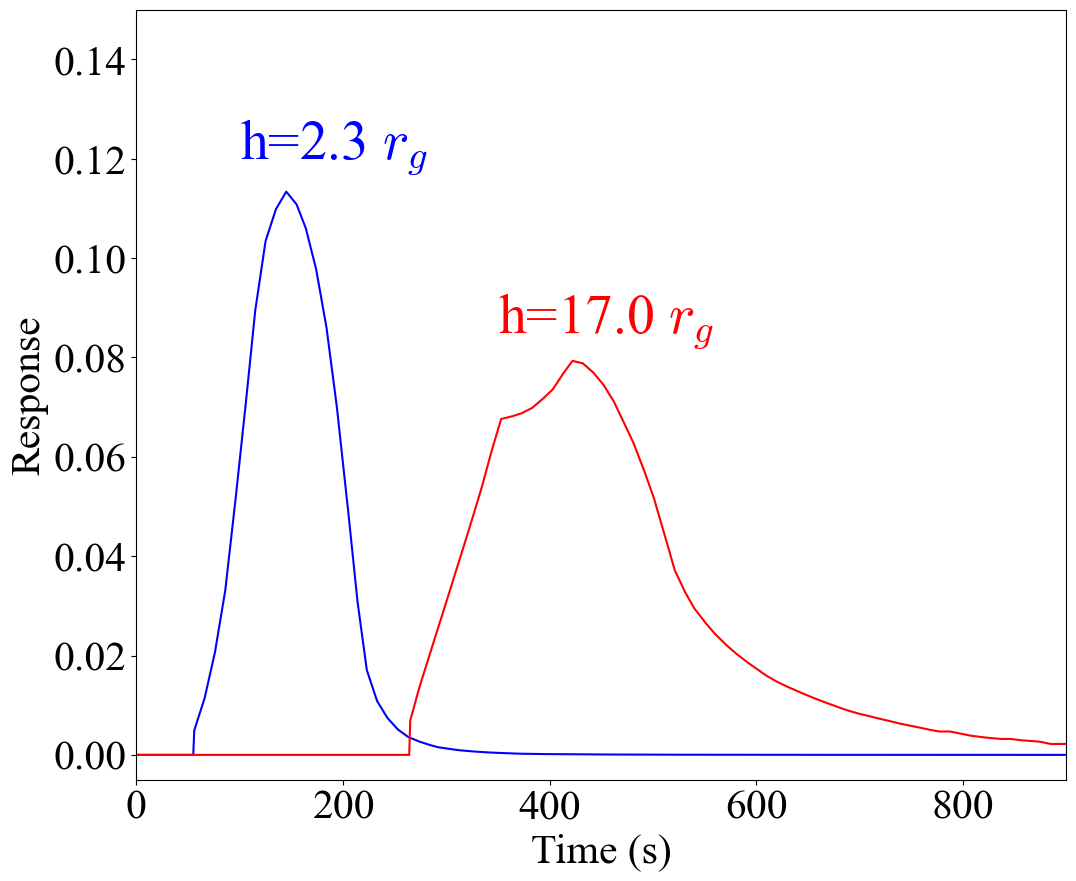} &
    \includegraphics[width=0.33\textwidth]{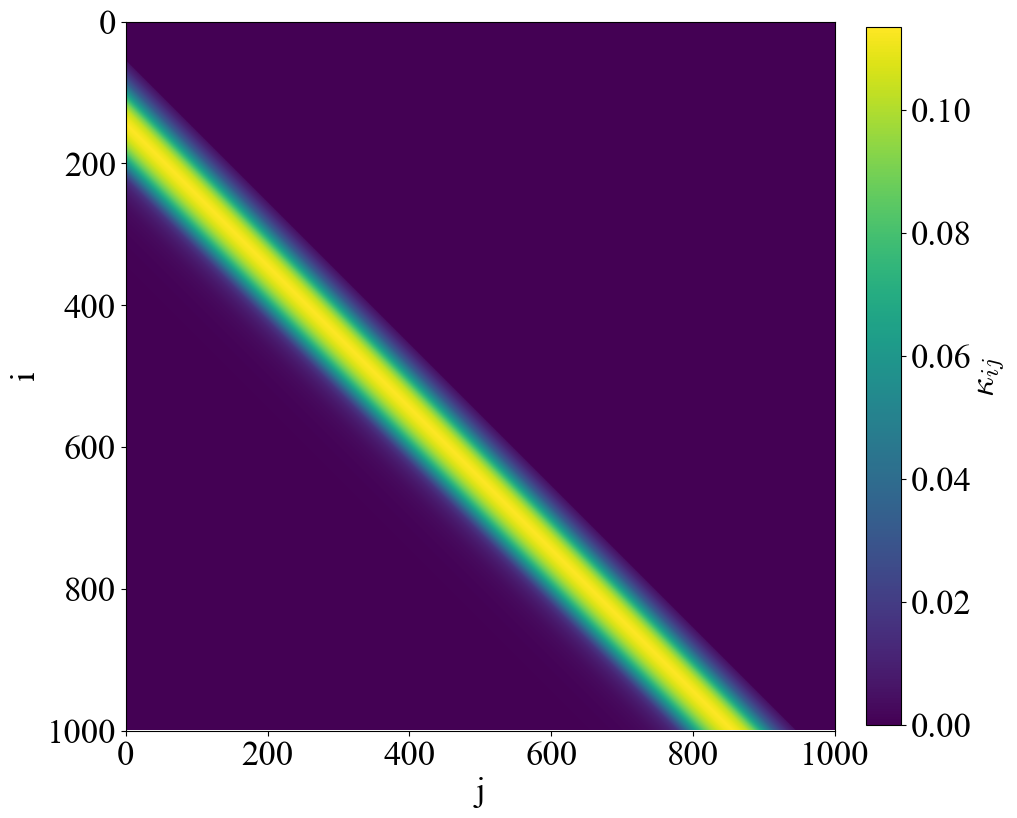} &
    \includegraphics[width=0.33\textwidth]{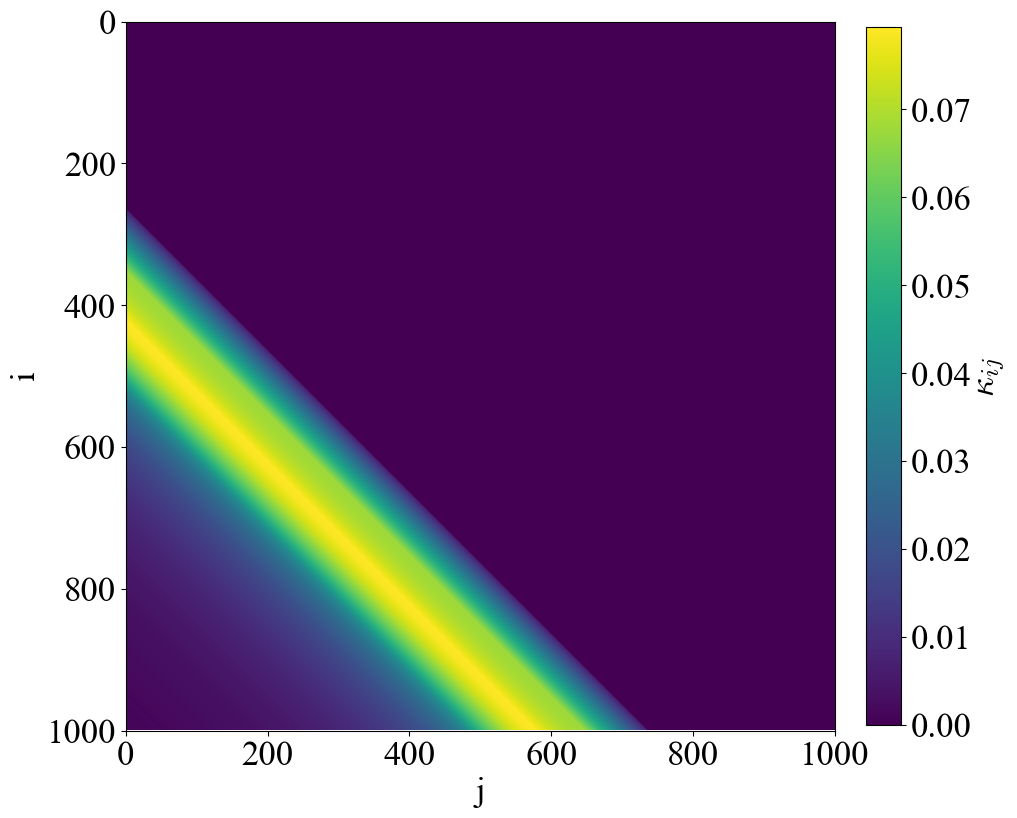}
  \end{tabular}
  \caption{\textit{Left panel}: Comparison between two response functions generated from the \textsc{KYNXILREV} code, adopting the lamp-post corona model with heights $h=2.3\,r_{\rm g}$ and $17\,r_{\rm g}$, where $r_{\rm g}$ is the gravitational radius. \textit{Center and right panels}: Two-dimensional matrix form of the response kernels with $h=2.3\,r_{\rm g}$ and $17\,r_{\rm g}$, respectively, presented by the color map. The conversion of the response function to the matrix form is described in Sec. \ref{sol_response}.}
    \label{fig:responses_heatmap}
\end{figure*}

This section is dedicated to finding the solution for the response kernel in the one-process (Eq.~\ref{eq:lc}) and the two-processes (Eq.~\ref{eq:lc_2p}) light curve equations. The central interests are how well the solved response kernels converge to the true ones and how robust the solutions are against the variation of the prefactors and the inclusion of the Gaussian noises. We choose the traditional lamp-post model for the corona geometry, generated using \textsc{KYNXILREV}, and the responses for the corona heights $h=2.3\,r_{\rm g}$ and $17\,r_{\rm g}$, where $r_{g}$ is the gravitational radius, are exemplified in Fig. \ref{fig:responses_heatmap}. The corresponding $\kappa_{ij}$ kernels, the form we proposed for the extraction method, are also illustrated in a color map alongside. As the corona height increases, the response shifts toward the bottom-left corner and becomes wider, in coherence with the response geometry in the temporal domain as the response function shifts toward the longer timescale. In this section, the response function for the corona height of $2.3\,r_{\rm g}$ is served as the ground truth.

\subsection{Effect of bin size}      
\label{ssec:1p_sol_bin}

The original light curves that span $100$ ks with the resolution of $1$ s yields the non-affordable numerical cost of the minimization problem, so the reduction of the data points is necessary. We start the presentation of the workflow performance with examining how the data reduction affects the robustness of solved one-process response kernel. We adopt the binning method of width $\Delta t$ in which all data points in the bin are averaged and centered. The choice of the bin size can be crucial as it reflects the resolution of the solved response kernel since we formulate the light curve equations into the discrete matrix form. We should be careful of the fine information of the light curves from the binning process. However, we assume at the first place that if $\Delta t$ is much smaller than the characteristic temporal features of the response function such as the offset, the centroid, or the width, the information of the kernels is not significantly lost. 

\begin{figure}
    \centering
    \includegraphics[width=0.5\textwidth]{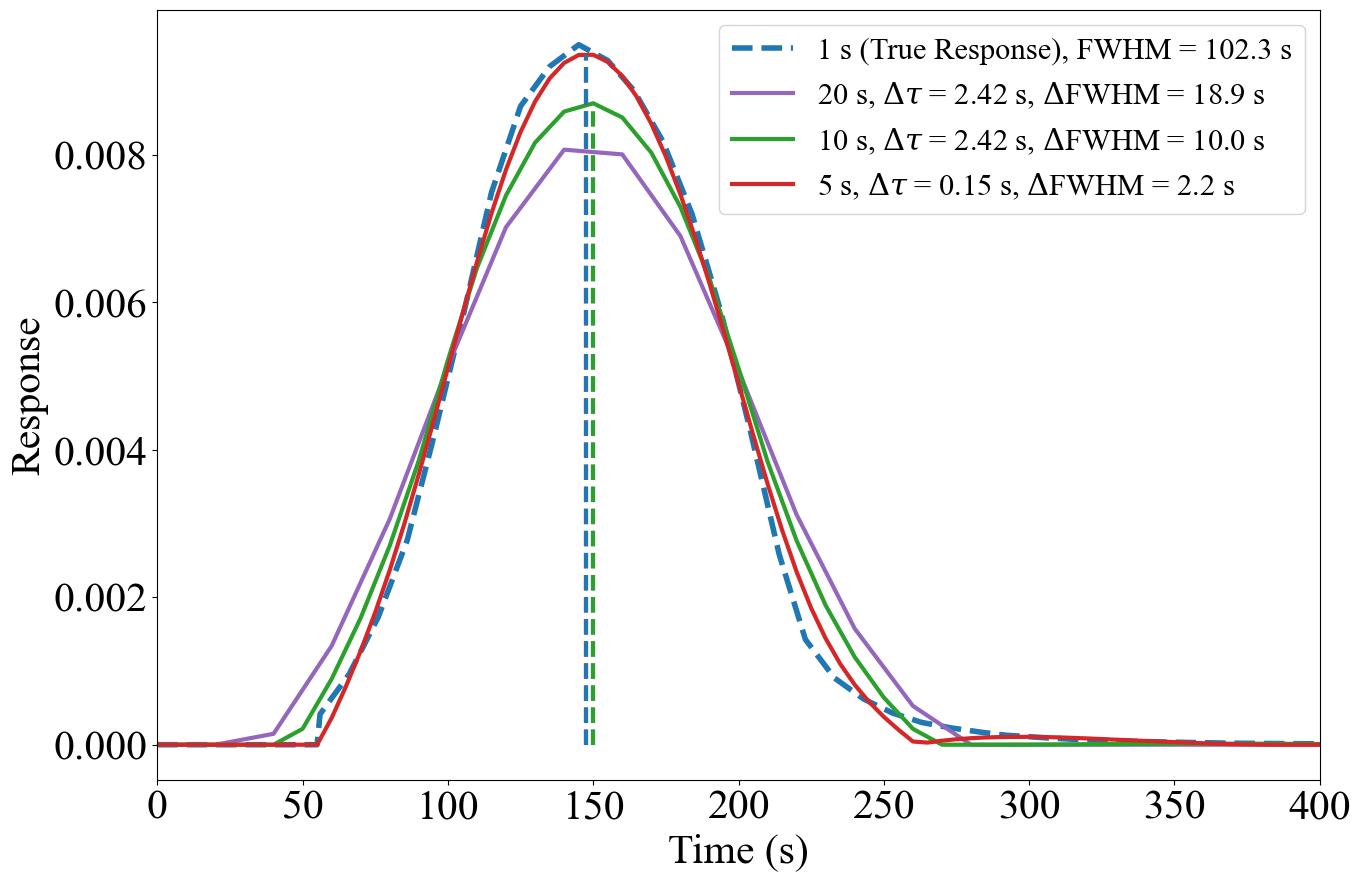}
    \caption{Solved reverberation responses of the lamp-post corona at $h=2.3\,r_{\rm g}$ for different $\Delta t=5, 10$ and $20$ s. The true response function is provided therein. The centroid shift $\Delta\tau$ and the FWHM deviation $\Delta\tau$ for each solved response function with respect to the true response are also indicated.}
    \label{fig:1P_bin_size_effect}
\end{figure}

Solved response functions for different $\Delta t$ in comparison with the ground truth are shown in Fig.~\ref{fig:1P_bin_size_effect}. 
We further evaluate the closeness between the true and the solved response functions by the centroid shift $\Delta \tau$. A centroid of a response function corresponds to the weighted mean of the time from the part above $80\%$ of the maximum, in order to neglect the near-zero response. 
In addition, we measure the difference between the full width at the half maximum ($\Delta\text{FWHM}$) of the solved and the true kernels. Both indicators are provided in the figure.
The solved kernel agrees strikingly well with the true kernel if $\Delta t=5$ s, yielding lowest $\Delta \tau$ and $\Delta\text{FWHM}$. For a bigger $\Delta t$, the solved kernels deviate from the true one with larger $\Delta\tau$ and $\Delta\text{FWHM}$, but the two shifts are still small compared with the corresponding characteristic timescales. The deviation of the centroid shift can be translated into the time lag as follows. The time lag between the two kernels of $1 \ r_{g}$ of height apart, supposedly the lamp-post height, is equal to $20 \ \text{s}$. Regarding the centroid shifts from the three bin sizes, the offset of the solved centroid from the ground truth can be translated to the time lag well below $20$ s. It is suggested from Fig.~\ref{fig:1P_bin_size_effect} that a bin size of $5$ s is an appropriate choice for further analysis without significant loss of detail of the response kernel.

\subsection{Test of robustness from the variation of prefactors for the one-process light curve}
\label{ssec:1p_sol_pref}

As we demonstrated in the previous section, the correct response function could be retrieved by our method provided that the input prefactors in the light curve equations are specified exactly to those used to generate the light curves.
In this section, we will test the sensitivity of the solutions to the variation of the prefactors from the right values in the cases involving one response kernel. Our scopes are, firstly, to examine how the solved response function alters by the variation of the prefactors around the right ones, which are unknown when dealing with the observed light curves. Technically, the workflow to obtain the optimized response kernel from a suite of light curves is processable for any chosen set of prefactors, although they are not fixed to the right ones. As the second objective, a suitable indicator for the closeness of the solved response kernels from various combinations of the prefactors to the true one shall be established. It is worth reminded that the available information of the AGN system in only the light curves, thus the indicator shall be based on that information.

We remind that the true prefactors in this investigation are $b_s=2.33$, $R_s=1.0$, $b_h=0.63$, and $R_h=0.5$. The variations of $b_s$ and $b_h$ are expressed in the units of the true prefactors for the direct components specified, namely, $b_{s0}$ and $b_{h0}$. For instance, the solution for the choice of $1.025b_{s0}$ and $1.025b_{h0}$ indicates that the solved response kernel is specific to the values of $b_{s}$ and $b_{h}$ equal to $1.025$ times of the right values while the reverberation prefactors are obtained from the flux conservation constraints in each band, namely, $b_{i0}+R_{i0}=b_{i}+R_{i}$. The closeness between the true and the solved kernels is evaluated indirectly by the maximum of the normalized root mean squared error (RMSE) $S_{h,\hat{h}}$ between the true (or the observed) light curves $h_l,i$ and the light curves synthesized from the optimized kernel $\hat{h}_l,i$, defined as 
\begin{equation}
    S_{h,\hat{h}} = \max \Biggl( \frac{\text{RMSE}(h_{l},i,\,\hat{h}_l,i)}{\mathbb{E}[h_l,i]} \Biggr) ,\; l \in \{s, h\}, 
    \label{eq:normalized_RMSE}
\end{equation}
where $\mathbb{E}[h_l,i]$ denotes the expectation values of the photon count rate, which is set to be the mean for our process. A smaller value means a less deviation of photon count from the true light curve, which can be inferred that the solved response function is closer to the true one.

\begin{figure*}
  \centering
  \begin{tabular}{cc}
    \includegraphics[width=0.4\textwidth]{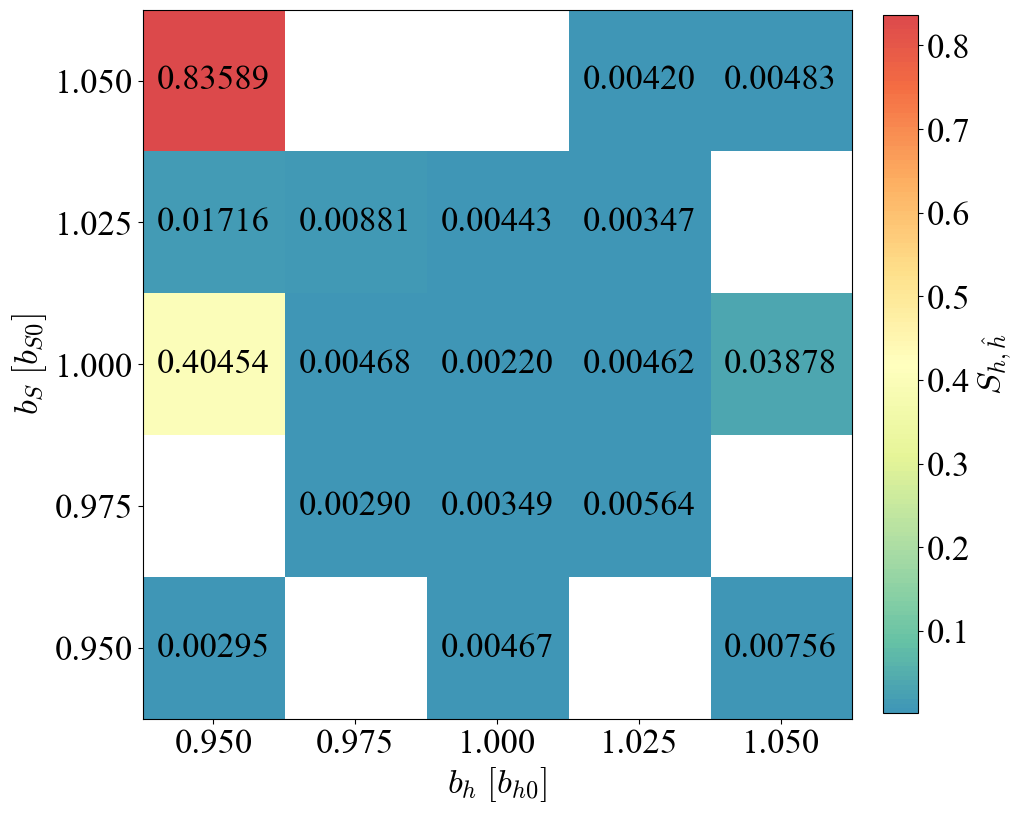} &
    \includegraphics[width=0.48\textwidth]{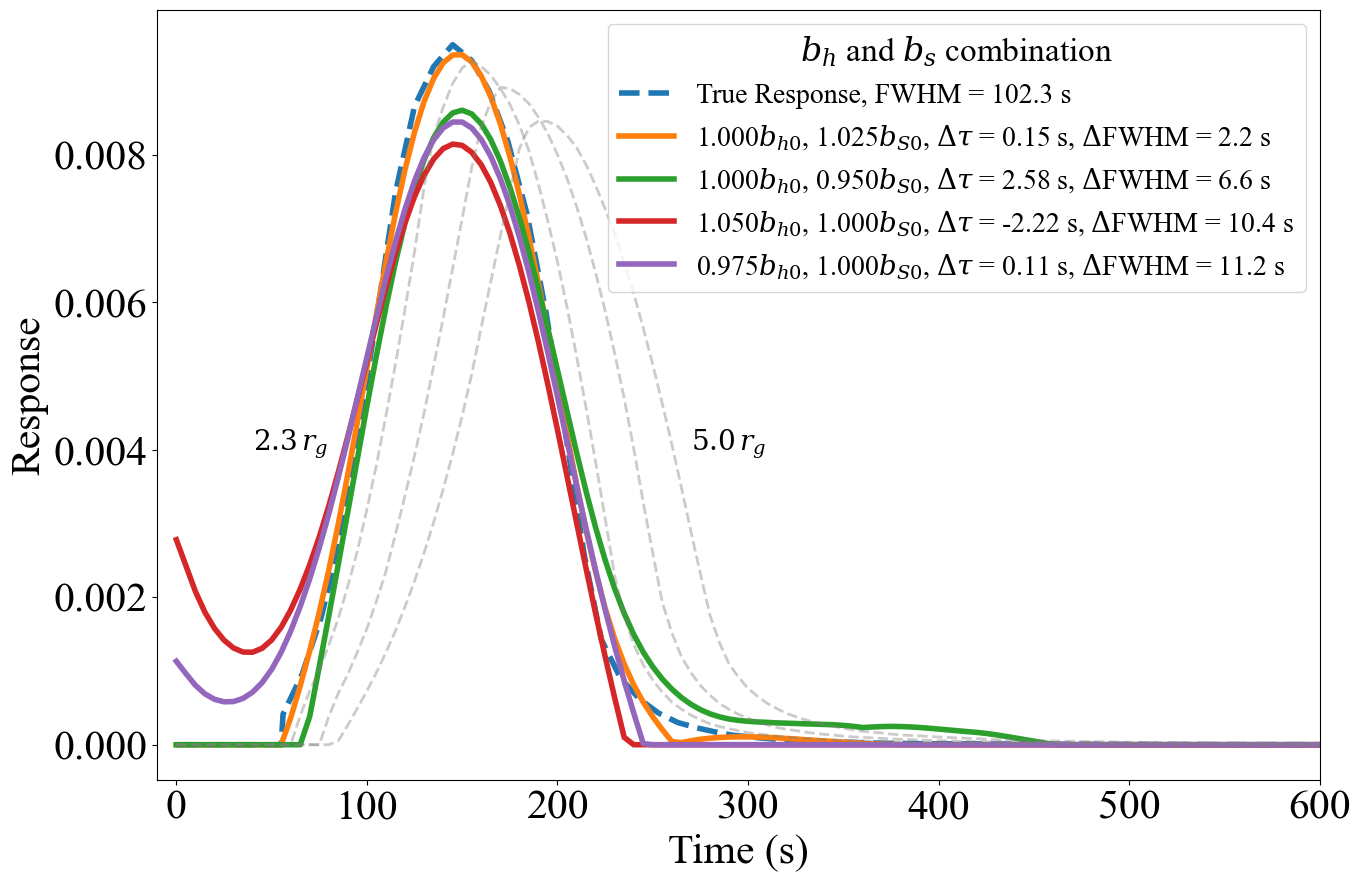}
    \end{tabular}
    \caption{\textit{Left panel}: Heatmap of $S_{h,\hat{h}}$ between true and regenerated light curves, expressed as a function of $b_{h}$ and $b_{s}$ for all tested cases. \textit{Right panel}: Comparison of some numerically solved response functions with the true response function (thick blue dashed line). Kernels of neighboring heights are included for comparison (gray dashed line). The heights of the dashed kernels (from left to right) are $3, 4$ and $5 \ r_{g}$. Centroid shift $\Delta \tau$ and FWHM deviation $\Delta\text{FWHM}$ are computed with respect to the true response and provided.} 
    \label{fig:1p_variation}
\end{figure*}

The maxima of normalized RSME for all tested variations within $5\%$ of the true prefactors is summarized in a color map on the left panel of Fig. \ref{fig:1p_variation} while the geometries of solved kernels for some selected cases are depicted on the right panel, with the choices of the prefactors indicated therein. We additionally provide the similarity metrics including $\Delta\tau$ and $\Delta\text{FWHM}$. As we expect from setting $b_{s}$ and $b_{h}$ to the correct values, the recovered kernel yields the lowest $S_{h,\hat{h}}$. The displacement within the $0.025b_{i0}$ vicinity in the color map leads to a higher $S_{h,\hat{h}}$ but it is not as high as those with either $b_{s}$ or $b_{h}$ varied by $5\%$, on average. 
We note a remarkably higher $S_{h,\hat{h}}$ when $b_{h}$ is altered by $\pm 5\%$ compared with the variation of $b_{s}$ by the same fraction. This implies that the kernel is more sensitive to the variation of the prefactors in the hard band. 
This can be explained that the hard-band light curve reverberation signal is more significant relative the the direct signal, as evaluated by the ratio between the reverberation to the direct factors, than the soft-band counterpart is. Thus, the variation of the hard-band reverberation prefactor affects more the kernel retrieval process. However, this is the technical issue specific to our choice of the light curve partitions. It does not signify that the hard-band light curve is more difficult to be handled. 
The retrieved kernel geometries for some selected sets of $b_{s}$ and $b_{h}$ are in accordance with the color map of $S_{h,\hat{h}}$. It can be seen that varying $b_{h}$ leads to a more noticeable discrepancy than varying $b_{s}$ by the same fraction. More specifically, the kernel can still be recovered almost entirely, with a minor mismatch at the right tail, if we increase $b_{s}$ by $2.5\%$ from $b_{s0}$, i.e., the orange curve, while decreasing $b_{h}$ from $b_{h0}$ by the same percentage, corresponding to the purple line, gives rise to a kernel with more notable discrepancy on the left side. Considering the two other cases, the green one with $b_{s}$ changed by $0.05b_{s0}$ still exhibits a reasonable agreement with the ground truth. Doing so in the hard band yields the kernel with the detail of the onset completely lost. However, if we recompile all kernels in the vicinities of the ground truth together, most of the features of the kernel such as the onset, the centroid, and the width are reasonably well recovered from these response matrices. 
The variation of $S_{h,\hat{h}}$ around the true prefactors in the color map is in coherence with the resulting $\Delta\tau$ and $\Delta\text{FWHM}$ in the right panel. More specifically, the average height of all solved kernels, estimated using the similarity metric with the kernel library, is equal to $2.7 \pm 0.4\,r_g$. The average height lies within $2.3$ and $3 \ r_{g}$ whose responses do not much differ visually as seen by the $2.3$ and $3 \ r_{g}$ kernels.

Our workflow demonstrated that the solved response closest to the true one can be obtained when all prefactors are specified to those used for generating the light curves. When we investigate the real system, those values and the true response function underlying the generation of the observed light curves are unknown. Hence, the consideration of the closeness to the true response is not practical in reality. We address this limitation by the utility test of a parameter involving the normalized RMSE between the observed and the synthesized optimal light curves, i.e., $S_{h,\hat{h}}$, and the results suggest that the values of such indicator correlates with the closeness between the true and the solved kernels. Instead of the true response function from the observed light curves, we should aim for the response kernel with the lowest possible $S_{h,\hat{h}}$. As the true prefactors are also unknown, the grid search in the $b_{i}-R_{i}$ space shall be deployed in order to solve for the best-optimized kernels. Despite a finest possible grid, it is possible that the $b_{i}-R_{i}$ grid points miss the true minimum of $S_{h,\hat{h}}$.
 
Nonetheless, the results we presented so far permit some flexibilities for the deployment to the real system. We demonstrated that the guessed prefactors are not necessarily to be exactly the right ones in order to recover the kernel reasonably close to the answer. The color map shows that even if the prefactors differ by up to $5\%$ from their true values, the recovered kernels remain nearly identical to the original. Therefore, when performing the grid search over these parameters, a grid resolution of about $5\%$, i.e., a step size of $0.05$ for quantities near unity, is adequate. In practice, using a coarser grid does not cost much computational time but increases the risk of missing the true values of $b_{s}$ and $b_{h}$, whereas using a fine grid since the start is numerically costly. Our results suggest that it is not necessary to employ a finest possible grid resolution since the start. A finer grid search can be localized at the points on a coarse grid, whose resolution can be fixed to $0.05$, at which the retrieved response function manifests the interpretable form.

\subsection{Inclusion of Gaussian random noise}
\label{ssec:1p_sol_noise}

Meanwhile, the robustness of solutions after the inclusion of the Gaussian noise can be examined. We use the noise model described in Sec. \ref{formula_response} which incorporates the zero-mean unit variance Gaussian function as a noise generator, and the noise of the designated signal-to-noise ratio is obtained via the factor $\alpha$ before the Gaussian function. 
In addition to directly solving for the response kernel from the noisy signals, we apply a bilateral filter, adopted from \citet{osti_1165751} using the source code developed by~\citet{suzuki_tsfilter}, to the noisy signals before the binning process to investigate if the results improve by such filtering. The filter is set to have the standard deviation in the time domain equal to $10$ s and the standard deviation in the count rate domain of $\sigma_{h}$ where $\sigma_{h}$ is the standard deviation of the observed count rate.

\begin{figure*}
  \centering
  \begin{tabular}{cc}
    \includegraphics[width=0.45\textwidth]{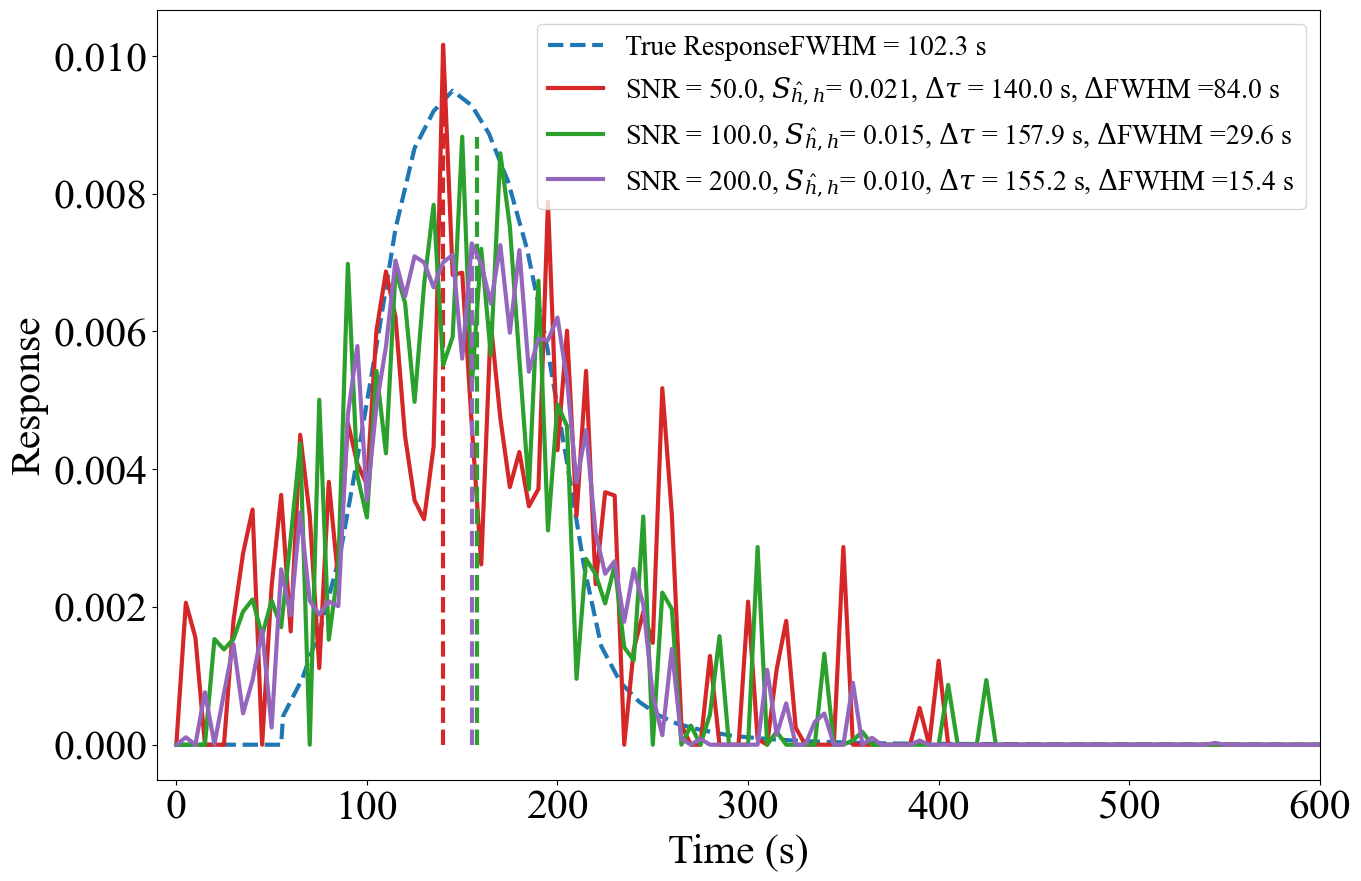} &
    \includegraphics[width=0.45\textwidth]{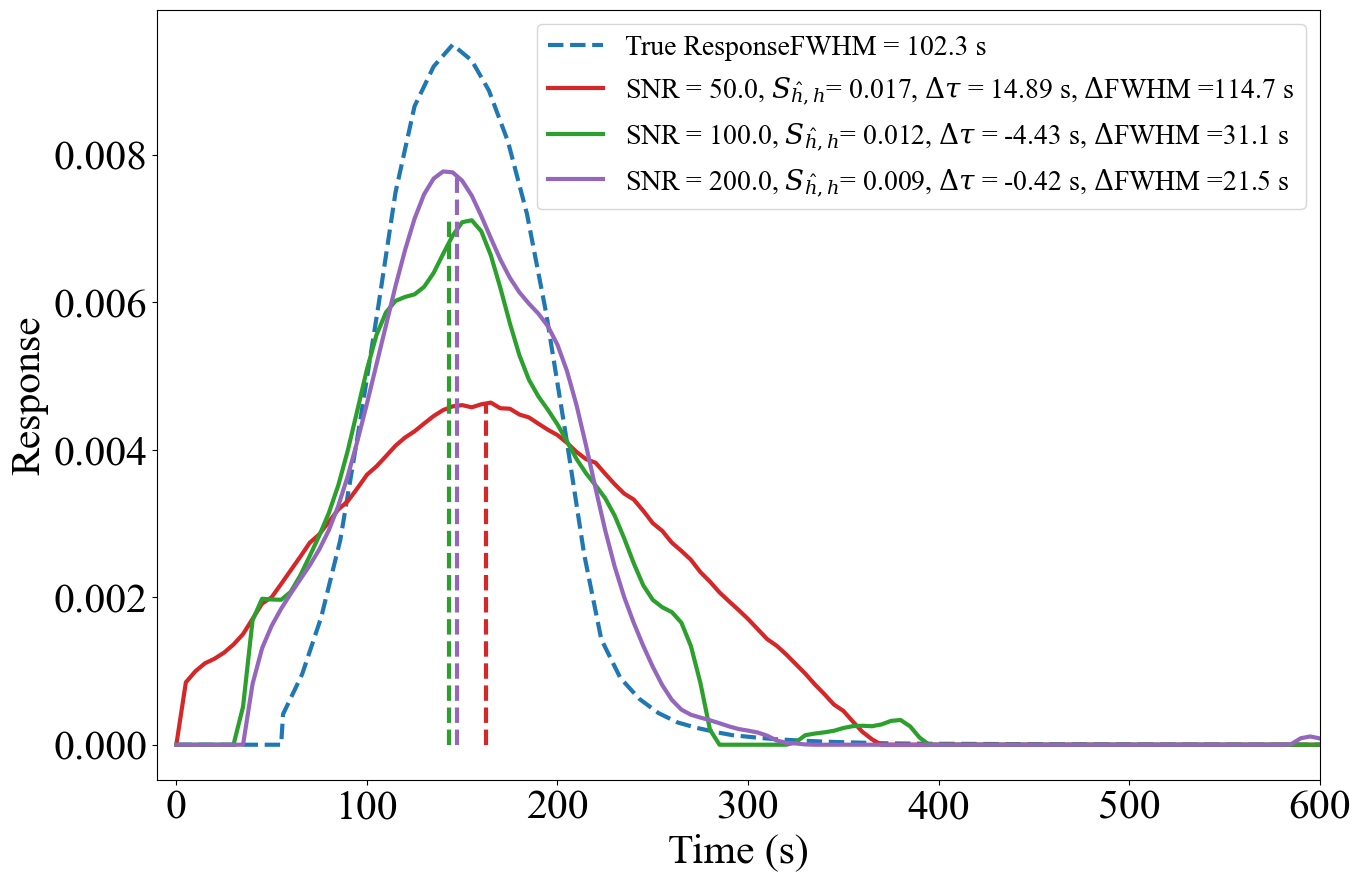}
  \end{tabular}
  \caption{Comparison of solved response functions in cases of SNR $=50, 100$, and $200$ with the ground truth. The left panel exhibits the solved kernels from the noisy signals, while the right panel is for the filtered signals using a bilateral filter prior to the optimization. The maximum normalized RMSE, $\Delta\tau$ , and $\Delta\text{FWHM}$ with respect to the true response are indicated therein.} 
    \label{fig:1P_Noise}
\end{figure*}

Effect from the noises to the accuracy and the performance of the filtering are summarized in Fig.~\ref{fig:1P_Noise} which demonstrates the solved kernels from the noisy and the filtered light curves, with $\Delta\tau$ and $\Delta\text{FWHM}$ indicated therein. It is evident that applying a bilateral filter improves the smoothness of the solved kernels. In terms of the SNR, a more elevated SNR yields a better agreement with the true response, as evaluated by the visual inspection and both indicators. The plot suggests that, with a proper denoising process, the SNR above $100$ yields the solved kernel reasonably close to the true kernel. However, the kernel obtained from the SNR$=50$ still renders a centroid, which is one of the key features of the response function, reasonably close to that of the ground truth. The normalized RMSE slightly improves after we apply the filter. These results underline the importance of the development of the denoising method along with the workflow. This topic could be considered in the future work.

\subsection{Performance of the method for the two-process light curves}
\label{ssec:2p_solutions}

In continuity with the tests of the accuracy of the solved response function from the one-process light curves as well as its robustness to artifacts, we investigate here if that method is able to solve for the two separate response functions. The two chosen kernels are the reverberation response function, as adopted in the one-process case, that dominates the short timescale at $\sim 100 \ \text{s}$ and the propagation response function represented by the top-hat function centered at $10^{4} \text{s}$ with the width of $10^{4} \ \text{s}$. The latter response represents the long-timescale process. Of interest is how well the two response kernels are retrieved and how the variation of prefactors affects the solutions.

\begin{figure*}
    \centering
    \begin{tabular}{cc}
      \includegraphics[width=0.45\textwidth]{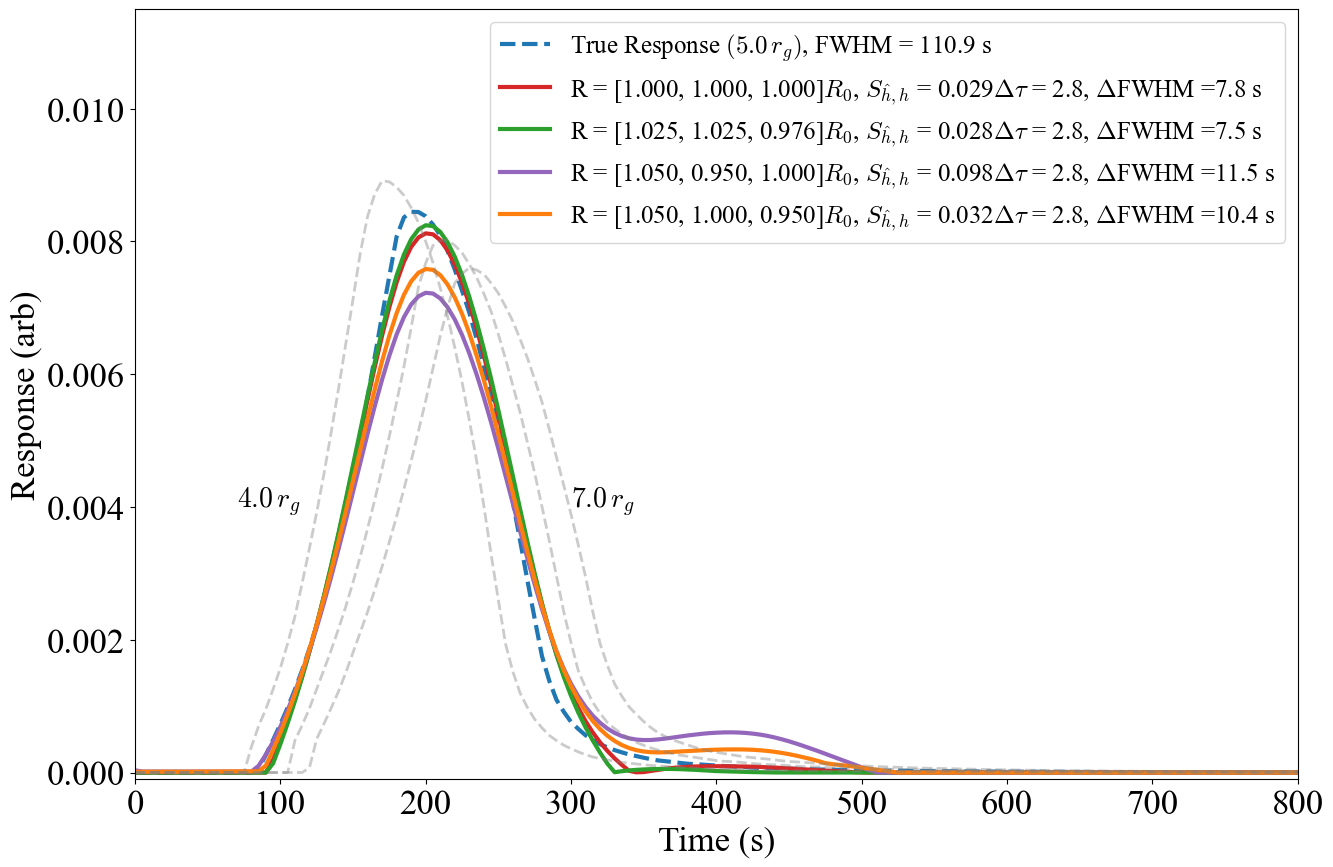} &
      \includegraphics[width=0.45\textwidth]{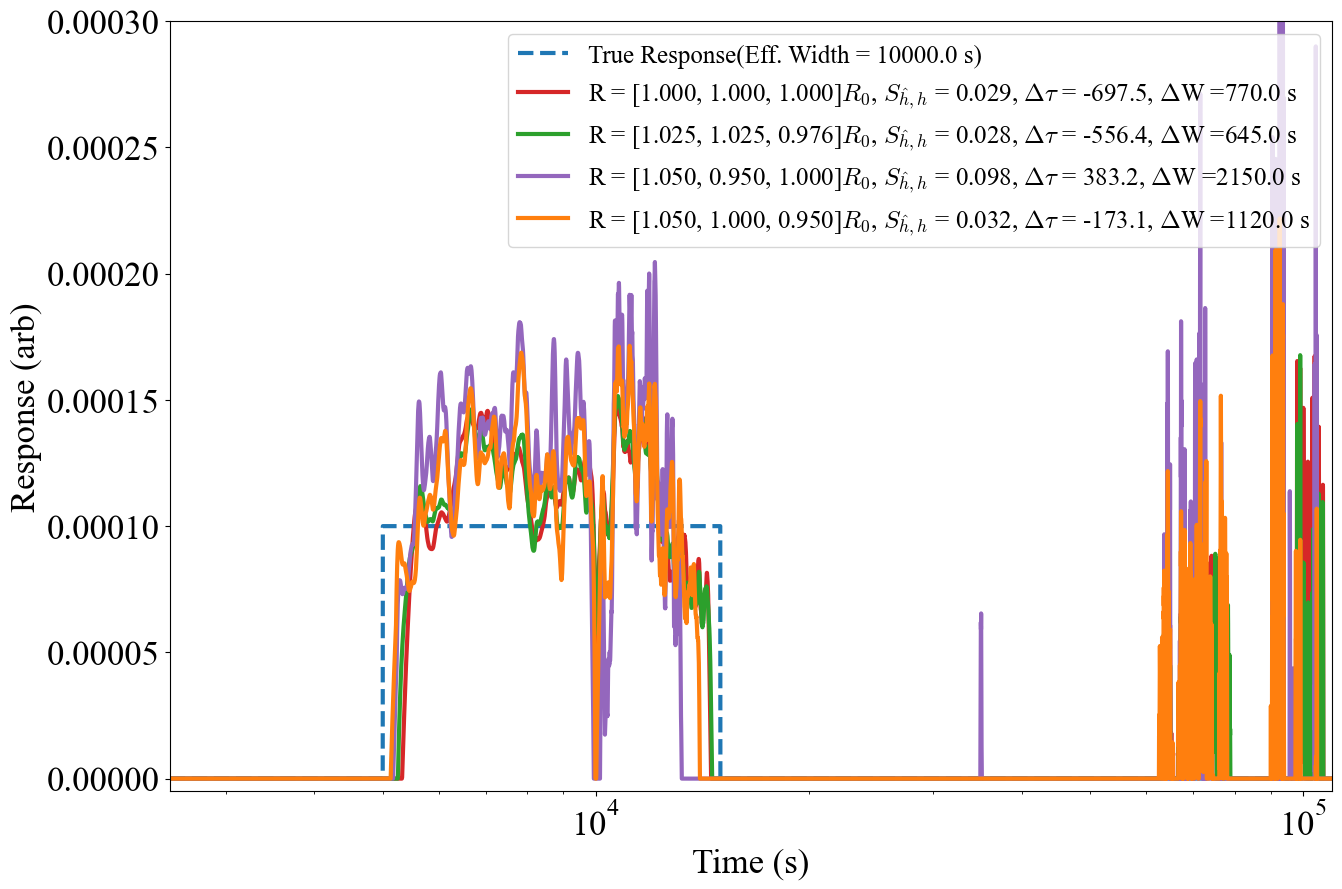}
    \end{tabular}
    \caption{Solved and true reverberation (left panel) and propagation (right panel) response functions with indicated varied prefactors. The change is presented by the fractions before the true reverberation prefactors in units of the true values $R_{0,i}$, put in order from the bands $1$ to $3$. The direct prefactors are unchanged while the change of the propagation prefactors is subject to the flux conservation. Similarity between solved and true kernels is presented by the centroid shift $\Delta\tau$, the FWHM deviation $\Delta\text{FWHM}$ (for the reverberation responses) and the effective width deviation $\Delta W$ (for the propagation responses). Reverberation kernels with the heights equal to $4, 6$ and $7 \ r_{g}$ are plotted in the gray dashed lines from left to right.}
    \label{fig:2P_combination_response}
\end{figure*}

We test numerous combinations of the prefactors within $\pm 5\%$ from the true values and some selected cases are exhibited in Fig. \ref{fig:2P_combination_response}, with $S_{h,\hat{h}}$, $\Delta \tau$ and $\Delta\text{FWHM}$, exclusively for the reverberation kernel. In place of $\Delta\text{FWHM}$, the difference in the effective width between the solved and the true kernels $\Delta W$ is evaluated and indicated alongside $\Delta\tau$ in the figure. The effective width is calculated simply by the time window between the non-zero bins at the two sides. The visualization of the results of all combinations involving the three light curves, each of which is linearly constituted from the three separate terms, is impossible as in the previous section but we do find a generic alteration of the solved kernels with the magnitude of $S_{h,\hat{h}}$ and all offset indicators. We present only cases with the direct fractions $b_{i}$ unchanged while we vary the reverberation fractions of the bands $i=1,2$ and $3$ by the indicated fractions and the propagation fractions are determined by the flux conservation constraint. 

With our $100$ ks simulated signal, the solution robustness is reasonably retained within $5\%$ of the parametric variation, as we concluded from the one-process study. As we expect, the reverberation kernel obtained from the true prefactors, i.e., the red curve, is the one among the kernels with the best agreement, and both kernels are cleanly separated due to the significant difference in timescale. Only small deviations on the right tails are observed. Cross matching analysis indicates that the average height is equal to $5.5 \pm 1.1\,r_g$, with respect to its ground truth scale height at $5.0\,r_g$. The kernels of adjacent indicated heights are provided for comparison. A response function at a long timescale is more poorly retrieved because, referring to the structure of the response matrix $\kappa_{ij}$ in Eq. (\ref{eq:kappa_mat}), the propagation kernel elements which lay at the timescale of $\sim 10^{4}$ s, are $100$ times less in number than the reverberation kernel elements, which has the characteristic timescale of $\sim 100$ s. Therefore, the propagation kernel has less significance in the loss function than the reverberation kernel does by $\sim 100$ times. 

Since the loss is determined from the two kernels of equal weights, the errors, which is more prominent in a kernel at a longer timescale, can be propagated between the convoluted signals in solving for the most-optimized ones. It leads to a slight deviation of the reverberation kernel although it is solved from the exact values for the light curve generation. 
The agreement of the two recovered kernels with the ground truths and $S_{h,\hat{h}}$ of this case are comparable to those of the green case where the reverberation fractions are deviated by $2.5\%$ at most. It is attributed to the error originating from the long-timescale kernel which has a poorer agreement. However, the cases within $2.5\%$ of the variations are those among the cases with the best agreement and the least $S_{h,\hat{h}}$.
The variation by $5\%$ yields a higher $S_{h,\hat{h}}$, on average, and a more deviation but the reverberation kernel's overall forms are still acceptably close to the solution, whilst the agreement for the propagation kernel is not improved nor worsened. 
As with the one-process counterpart, the norm of $S_{h,\hat{h}}$ which is estimated directly from the light curves, varies in line with the three indicators of the geometrical shifts.
The test on the two-process workflow conforms with the numerical test for the one-process case such that if the grid search is to be applied for the search of the kernels, the grid resolution of $\sim 0.05$ suffices in locating the least $S_{h,\hat{h}}$ point. 

A fewer number of elements in the response kernels at large $t$ is also the explanation of the emergence of the noises at $10^{5}$ s, which even has lower significance in the loss function during the optimization than the propagation kernel does. The noise level is higher when we displace more from the true prefactors. In order to compensate a lower weight in the loss function for the long-timescale kernel, the time-weighted loss function might be of use.

\section{Discussion and conclusion}
\label{sec:conclu}

We introduce an alternative method to extract the X-ray reverberation response functions from the multi-band X-ray light curves. We rely on the numerical optimization procedure based on PyTorch \citep{Pytorch} for the workflow. It determines the optimal response function corresponding to that generating the light curves closest possible to the original light curves. 
To solve for the $n$ underlying response functions, $n+1$ different bands of the light curves at the same time span are required. We assume the light curve form of all bands to be the superposition of the direct and the convoluted processes originating from the same responses. 
It must be noted that the assumption of the identical driving signal might not be hold in the cases of the extended corona, the different emission region for each energy band, or the energy-dependent emissivity. Also, the reverberation response functions across X-ray bands can be energy-dependent \citep{reynolds_et_al_1999,Cackett2014,Chainakun2016,Epitropakis2016,Wilkins2016}. However, one may assume energy-independent responses when a single light curve is extracted over a sufficiently broad energy band such that the majority of Doppler- and gravitationally-shifted photons associated with a single feature in the reflection  spectrum (e.g., the broad Fe line ) are encompassed. The possible direction for a more flexible workflow incorporating the energy-dependent responses can be as follows. If we restrict to the current requirement, namely, a suite of $n+1$ bands for $n$ underlying responses, a known inter-relation between responses in different bands is necessary. Otherwise, more bands are required which make the optimization process more complicated.

The driving signal is solved at first based on the choices of the prefactors of each contribution before the optimization for the kernels is proceeded on. We have demonstrated the feasibility of this method up to $n=2$. Our optimization workflow is capable of solving for the response function with acceptable closeness to the real one and is proved tolerant to a certain degree of the physical/numerical artifacts. 
This method can be of use as a complementary method to the conventional one. While our investigation focuses on the lamp-post geometry, the method itself does not require any assumed geometry at the outset. By virtue of the flexibility our method provides, the solved kernel, as a solution of a valid set of prefactors, can be compared with the responses derived from the other traditional methods for the consistency. The lamp-post model is not the sole theoretical model in the model pool used to investigate the corona geometry, but there are also other choices such as the extended corona model \citep{Wilkins2016, Chainakun2019, Hancock2023}, the dual lamppost model \citep{Chainakun2017, Hancock2023}, the radially extended corona model \citep{Wilkins2016, Chainakun2019}, the warm corona model \citep{Kubota2018, Ursini2020, Xu2021, Ballantyne2024}, or the accretion disks with realistic geometric thickness \citep{Taylor2018}. Therefore, a flexible method capable of solving for the response function geometry without assuming its geometry beforehand could be a useful tool for a better understanding of the AGN system.

When applying this method to observed light curves, it should be reminded that the form of the true response functions and the prefactors underlying the generation of the observed light curves are unknown. This unknowing can be overcome by a grid search over the hyperparameter space of the prefactors, and the retrieved kernel thus represents the optimal kernel specific to the chosen set of the prefactors. As demonstrated in Sections \ref{ssec:1p_sol_pref} and \ref{ssec:2p_solutions}, the root mean square error (RMSE) relative to the the mean count rate (or the normalized RMSE) between the reconstructed and observed light curves correlates with the accuracy of the recovered response functions. While a coarser grid can significantly economize the computational cost, it carries the risk of overlooking the best-fitting and most physically meaningful solutions. In high-stakes analyses where accurate recovery of the response function geometry is essential, this trade-off must be carefully considered. Our accuracy test suggested that the grid resolution of $0.05$ for the prefactors suffices to inform the proximity to the least-RMSE kernel. A finer grid may additionally be applied locally.
Furthermore, the inclusion of physical constraints, such as requiring the recovered driving signal to be mostly positive, can further screen out some unphysical combinations of the prefactors from consideration. 
However, we did not thoroughly investigate the degeneracy of the solution arising at the other point far from the point of the true prefactors in the hyperparameter space as we displaced ourselves only $5\%$ from the true prefactors at most. The existence of the other response function yielding comparable root mean square error might render complexity to the workflow. 
Another remark from the two-process case is that, although we demonstrated a robust workflow that yielded cleanly separate solved kernels, the situation can be complicated if the second kernel is at the proximity of the first one. The reverberation kernels solved in the two-process workflow exhibited prominent tails that could be $300 \ \text{s}$ of width, due to the error propagation between kernels, so it can be inferred that the two-process workflow has a limitation for the two kernels that are at least $300$ s apart. It is possible that the error propagated between the two close kernels can render a solved kernel that is uninterpretable.

Gaussian noises with an adjustable signal-to-noise ratio (SNR) are added to the light curves in order to investigate how they effect the accuracy of the the solved response kernel. A high level of the noises can significantly alter the obtained geometry of the response kernel but, with a proper treatment of the noise, some important geometric features such as the centroid, the width, or the offset can be retrieved with acceptable accuracy. The agreement with the true response kernel improves when the SNR increases. More specifically, the closeness is reasonable when the SNR $\geq 100.0$. 
Although our SNR definition that directly compared the power of the signal, i.e., the count rate squared, to the noise variance may differ from the common choices for astrophysical signals, it can straightforwardly be translated to those systems. For instance, the square root of our SNR can be regarded as the ratio of the mean signal to the standard deviation of the noises. Our $\text{SNR}=100$ is equivalent to the signal strength equal to $10$ from that definition, which is obtainable by modern equipments (see, e.g., \citealt{oh_et_al_2018}). Otherwise, for the signals contaminated with the shot (or Poissonian) noises which self-consistently relate with the signal level, the corresponding SNR can be estimated by $S/\sqrt{S+\mathcal{N}_{t}}$, where $S$ is the total count and $\mathcal{N}_{t}$ is the total noise across the exposure time. Our $\text{SNR}=100$ with an average count rate of $1-3 \ \text{s}^{-1}$ across $100 \ \text{ks}$ yields the equivalent shot-noise-based SNR in the range of $300-500$, a level attainable from observations with a long exposure time~\citep{gallo_2006,Nakhonthong2024}.
Our investigation suggests that the improvement of the denoising method is equally important as the extraction method, and the improved denoising method may advance greatly the knowledge in this domain.

As demonstrated in Sec. \ref{ssec:2p_solutions} for the two-process light curves involving the short-timescale reverberating response function and the additional long-timescale propagation response function, as employed in~\citet{Alston2014} and \citet{Chainakun2023}, the timescale at which the response kernel is better retrieved is that of the former kernel, i.e., $\sim 100 \ \text{s}$, whereas the latter one at a long timescale, namely, $\sim 10^{4} \ \text{s}$, is retrieved with less accuracy. It is attributed to a lower number of elements in the response kernel, thus it is less significant in the optimization process when the two kernels are solved together. This suggests that the time span of the light curves has to be at least $10$ times of the timescale of the designated kernel. It turns out that a long exposure time benefits both the SNR level and the retrievability of the kernel at a designated timescale.
Alternatively, an appropriate weighting technique could be incorporated into the computation of the loss function. For example, the different numbers of kernel elements of the short- and long-timescale kernels could be balanced by the time-dependent weighting function incorporated into the loss function.

In comparison between our proposed method and our previous study~\citep{deesamutara2025extractingxrayreverberationresponse} which was based on the Variational Autoencoder (VAE) trained with simulated X-ray light curves, and the trained set was employed to predict the reverberating response function. The VAE prediction achieves lower time-wise computational costs while the ability to predict the response function is limited into a specified range on parametric spaces, i.e., the black hole mass, the accretion disk inclination, and the coronal scale height, assuming the lamp-post corona model. 
Albeit the high cost of the computational time, our numerical solver gains an advantage, compared to the VAE-based pipeline, that the optimization pipeline does not rely on a specific model. Instead, it provides flexibility for the other models, e.g., sandwich corona~\citep{Beloborodov_1999}, spherical corona~\citep{Chainakun2019} to be solved for. Secondly, our proposed method can solve multiple processes at once, but the numbers of the input light curve bands required increase with the number of processes we aim for, while the VAE network in~\cite{deesamutara2025extractingxrayreverberationresponse} only requires one band of light curve to predict the response function. We add a note to this point that the requirement of multi-band light curves can be directly fulfilled by the co-observation by current X-ray instruments such as the XMM-Newton and the NuSTAR observatories, or obtained post-observationally by further processing such as the division of full-band light curves into the multi-band components~\citep{Chainakun2016, Hancock2022}.

In applying our framework onto the real light curves, several other challenges still remain. 
For our first challenge, it is possible that the observed light curve consists of missing data or the unequally-spaced data points, while our method currently relies on well-sampled and uniformly-binned light curves. It requires appropriate treatment to resolve the data gaps, e.g., the interpolation methods, and/or the forward modeling such the Gaussian process regressors~\citep{Wilkins2019_GP, Lewin_2022, PozoNunez_GP}. 
Secondly, the observed light curve consists of the Poisson noises, the effect of which is not tested here. The realistic Poisson noises are more complicated because their variances vary in each bin in proportion with the discrete count rate, differing from the fixed-variance Gaussian noise model that we test here. The effect of these noises can be important in the high-energy bands, which typically yield a low count rate. As the hard-band X-ray light curve is an important ingredient in our workflow, the accuracy of the kernel retrieval can potentially be affected. 
A number of studies suggested that the advanced filtering techniques, e.g., the non local mean filters or the composite bilateral filters  \citep{osti_1165751}, or the signal reconstruction using an auto-correlation analysis \citep{prh1992} removed noises efficiently.
All of these attributes can potentially lead to the follow-up studies in the future. 

Lastly, the mathematical and optimization framework developed here could also serve a broader scope than the X-ray reverberation-mapping applications. In particular, it can be adopted to long-term optical/UV reverberation mapping in AGN and quasars, with LSST~\citep{Czerny_2023, PozoNunez2023, Panda2024} for examples, where one seeks to invert the convolution kernel between the UV-driving light curve and the longer-wavelength response. This approach aligns well with recent progress in optical continuum reverberation studies. For instance,~\cite{PozoNunez_2025} reported a direct measurement of the accretion disk size in a quasar via multi-band continuum-lag analysis, demonstrating the feasibility of continuum reverberation studies at cosmological distances. Extending the present framework to such regimes could therefore provide a powerful tool for mapping accretion disk structures across cosmic time.

\begin{acknowledgements}
This work is supported by (i) Suranaree University of Technology (SUT), (ii) Thailand Science Research and Innovation (TSRI), and (iii) National Science, Research and Innovation Fund (grant number 215538). P.C. thanks funding support from National Research Council of Thailand (NRCT) and Suranaree University of Technology, grant number N42A680156. F.P.N. gratefully acknowledges the generous and invaluable support of the Klaus Tschira Foundation. F.P.N. acknowledges funding from the European Research Council (ERC) under the European Union’s Horizon 2020 research and innovation program (grant agreement No. 951549). S.D. acknowledges the support of External Grants and Scholarships for Graduate Students (OROG) by the Institute of Research and Development, SUT (Academic year 2568).
\end{acknowledgements}

\section*{Data Availability}
This work is developed using \textsc{PyTorch} framework \citep{Pytorch}, alongside \textsc{Pandas} \citep{reback2020pandas}, and \textsc{Astropy} \citep{astropy:2013, astropy:2018, astropy:2022}. Source code used in this article is available at \url{https://github.com/dsanhnt/Response_Solver}.

\appendix
\section{Test of convergence to a simple solution}
\label{app:test_dirac_delta}

\begin{figure}
  \centering
  \begin{tabular}{cc}
    \includegraphics[width=0.45\textwidth]{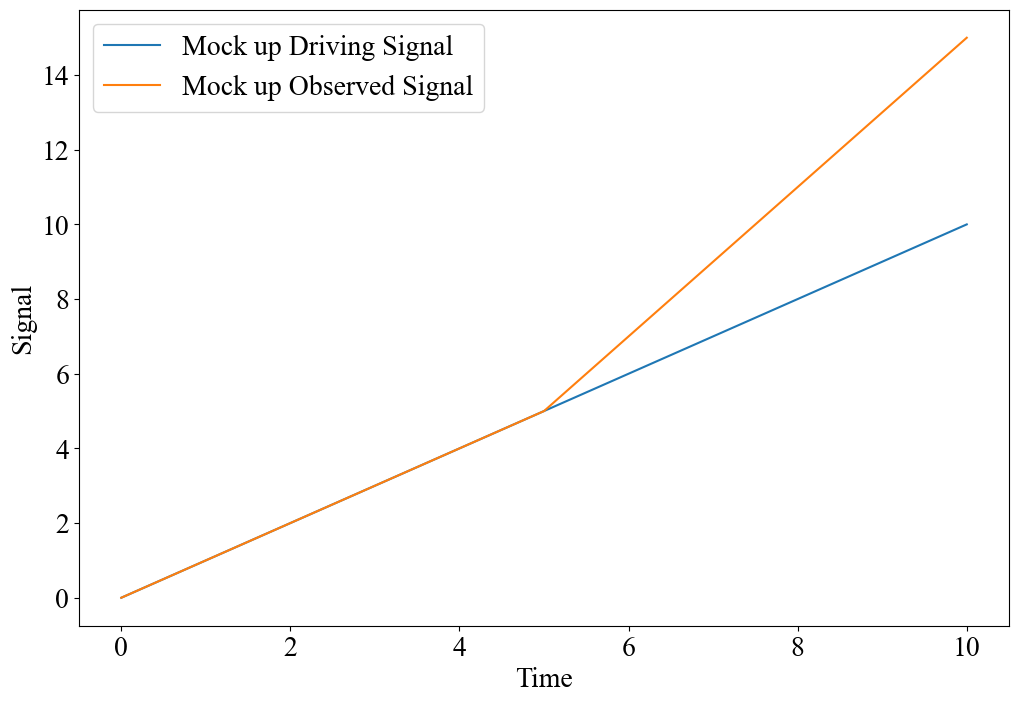} &
    \includegraphics[width=0.45\textwidth]{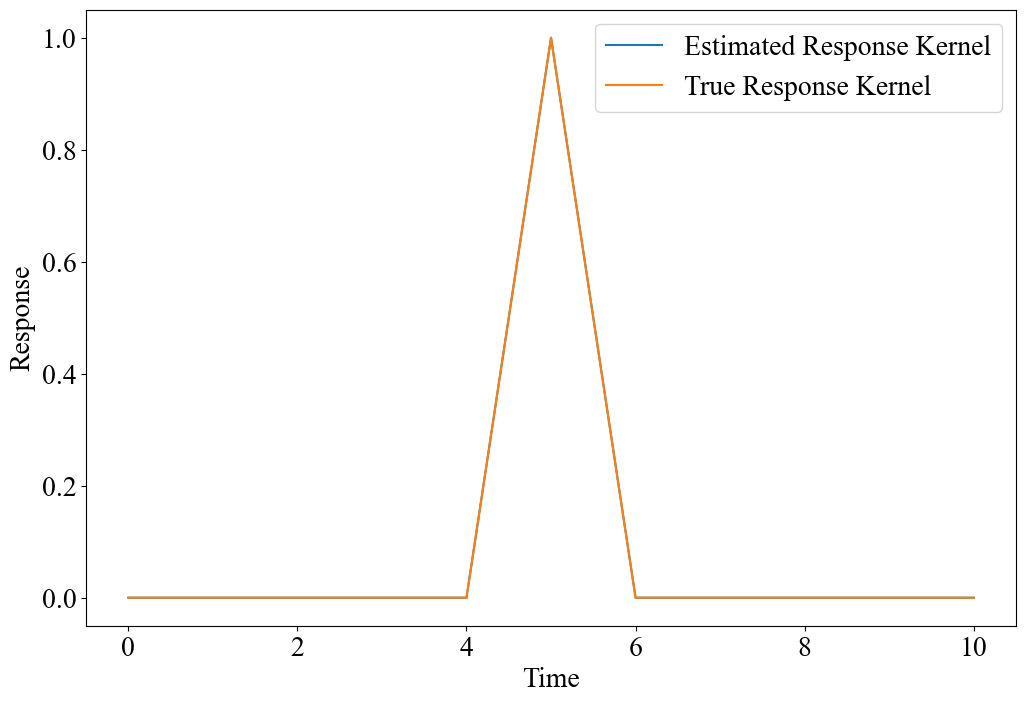}
   \end{tabular}
  \caption{\textit{Left panel}: The driving signal and the mocked-up simulated light curve used for the applicability test of the workflow. \textit{Right panel}: Solved response kernel compared with the true response kernel.}
   \label{fig:signals}
\end{figure}

To verify the method feasibility, as a proof of concept, we apply the developed optimization procedure to the driving signal in the linear function form $a(t)=t$ and the mocked-up single-banded observed signal is generated using Eq. (\ref{eq:linear_lc_iterative}) with Dirac delta response function ($\delta(t-5)$). As a result, the observed signal can be solved analytically and it reads 
\begin{equation}
   s_h(t) = 
   \begin{cases}
       b\,t & t < 5 \\
       (b+R_h)t - 5 R_h & t \geq 5
   \end{cases}
\end{equation}
We test our optimization workflow in the range of $t\in[0,10] s$, with the time step of $1$ s. The solved kernel, via the proposed optimization workflow, and the true kernel are depicted in in Fig.~\ref{fig:signals}. As shown in that figure, the optimization workflow can retrieve the response kernel as a step function centered at $t=5$, in agreement with the ground truth.

\bibliographystyle{aasjournal}

 \newcommand{\noop}[1]{}

\end{document}